\documentclass[12pt,fleqn]{article}

\usepackage{amsmath, amssymb,graphicx}

\newcommand{\CC}{\mathbb{C}}

\newcommand{\ZZ}{\mathbb{Z}}

\begin{document}


\thispagestyle{empty}
\renewcommand{\thefootnote}{\fnsymbol{footnote}}

{\hfill \parbox{3cm}{
       }}

\bigskip\bigskip\bigskip

\begin{center} \noindent \Large \bf

{\bf  Intersecting Branes, \\ Defect Conformal Field Theories\\ and Tensionless Strings}
\end{center}

\bigskip\bigskip\bigskip\bigskip

\centerline{ \normalsize \bf Neil R. Constable$\,^{a}$, Johanna
Erdmenger$\,^b$,
Zachary Guralnik$\,^{b}$ and
Ingo Kirsch$\,^{b}$\footnote[1]{\noindent \tt
constabl@lns.mit.edu, jke@physik.hu-berlin.de,
zack@physik.hu-berlin.de,  ik@physik.hu-berlin.de} }

\bigskip
\bigskip\bigskip

\bigskip\bigskip

\centerline{$^a$ \it
Center for Theoretical Physics and Laboratory for Nuclear Science    }
\centerline{ \it Massachusetts Institute of Technology}
\centerline{\it 77 Massachusetts Avenue}
\centerline{ \it Cambridge, {\rm MA}  02139, USA}
\bigskip
\centerline{$^b$ \it Institut f\"ur Physik}
\centerline{\it Humboldt-Universit\"at zu Berlin}
\centerline{\it Invalidenstra{\ss}e 110}
\centerline{\it D-10115 Berlin, Germany}
\bigskip\bigskip

\bigskip\bigskip

\renewcommand{\thefootnote}{\arabic{footnote}}

\centerline{\bf \small Abstract}
\medskip

{\small The defect conformal field theory describing intersecting
D3-branes at a $\CC^2/\ZZ_k$ orbifold is used to (de)construct the
theory of intersecting M5-branes, as well as M5-branes wrapping
the holomorphic curve $xy =c$.  The possibility of a 't Hooft
anomaly due to tensionless strings at the intersection is
discussed.  This note is based on a talk given by Zachary Guralnik at the 35th 
International Symposium Ahrenshoop on the Theory of Elementary Particles.}

\newpage


\section{Introduction}
\setcounter{equation}{0}

String theory has led to the discovery of many novel interacting
conformal theories.  Amongst these are the ``defect'' conformal
theories which describe intersecting branes at low energies. Such
theories were studied originally in
\cite{Sethi,GanorSethi,KapustinSethi} and more recently in the
context of AdS/CFT duality in \cite{Karchrandall1,Karchrandall2,
Karchrandall3,DeWolfe,EGK,Lee,Skenderis,Bachas,Quella,Mateos,CEGK}.

One of the more exotic examples of this type of theory is the one
which arises for intersecting M5-branes.  Because of the difficulty 
of constructing a non-abelian generalization
of a chiral two-form,  a Lagrangian description is lacking even for parallel M5-branes.  The dynamics of intersecting M5-branes is
even richer and less understood, due to the presence of
tensionless strings localized at the intersection \cite{Hanany}.
Until very recently,  the only concrete formulation of the theory was in
terms of a discrete light cone quantization using the matrix model
discussed in \cite{Kachru}.

In this note, we shall describe an alternate formulation,  in
which the theory of the M5-M5 intersection is obtained from a
limit of a defect conformal field theory with two-dimensional
$(4,0)$ supersymmetry \cite{CEGK2}. This formulation is a natural
extension of the ``(de)construction'' of the six-dimensional theory
of parallel M5-branes discussed in \cite{Arkani-Hamed,Csaki}. The
$(4,0)$ defect CFT describes the low energy limit of intersecting
D3-branes at a $\CC^2/\ZZ_k$ orbifold. In a suitable $k\rightarrow
\infty$ limit on the Higgs branch, the two extra dimensions of the
M5-M5 intersection are generated.

We shall begin by discussing the $(4,4)$ defect CFT which
describes D3-branes intersecting in flat space. This theory is the
low energy limit of the M5-M5 intersection on a torus. We shall
present the exact superpotential of the $(4,4)$ defect CFT in
$(2,2)$ superspace.  The resolution of the intersection to the
holomorphic curve $xy=c$ can be seen explicitly from solutions of
the F-flatness conditions. Upon orbifolding to a $(4,0)$ theory
and taking the appropriate limit, all the degrees of freedom with
momentum on the torus are generated.

In $(2,0)$ superspace, the analogue of a superpotential for the
$(4,0)$ defect CFT is readily obtained from the  superpotential
for the $(4,4)$ defect CFT.  At the appropriate point on the Higgs
branch, this superpotential can be interpreted as lattice kinetic
term in a compact direction, generating a Kaluza-Klein spectrum in
the $k\rightarrow\infty$ limit.  Via the S-duality of the model, a 
Kaluza-Klein spectrum on a
two-torus is generated. In the
continuum limit, the two-dimensional fields localized at the
intersection correspond to tensionless strings propagating in the
four common dimensions of the M5-M5 intersection.

The $(4,0)$ defect CFT has a $SU(2)_L$ R-symmetry. In the
continuum limit,  we shall argue that this becomes the $SU(2)$
R-symmetry of the M5-M5 intersection,  which preserves
four-dimensional ${\cal N} =2$ supersymmetry.  The $SU(2)_L$
R-symmetry exhibits a 't Hooft anomaly.  In the continuum limit,
this may indicate an anomaly in the $SU(2)$ R-symmetry due to
tensionless strings. While local $SU(2)$ anomalies do not arise in
local four-dimensional quantum field theories,  the possibility is
not excluded for a four-dimensional theory of tensionless strings.

\section{The D3-D3 intersection in flat space}
Consider a stack of $N$ parallel D3-branes in the directions
$0123$ intersecting an orthogonal stack of $N^{\prime}$
D3$^{\prime}$-branes in the directions $0145$. The action has the
form $S= S_{\rm D3} + S_{\rm D3^{\prime}} + S_{\rm
D3-D3^{\prime}}$. The components $S_{\rm D3}$ and $S_{\rm D3}$
each correspond to a four-dimensional ${\cal N} =4$ theory. The
term $S_{\rm D3-D3^{\prime}}$ contains couplings to a
two-dimensional $(4,4)$ hypermultiplet, leaving only $(4,4)$
supersymmetry unbroken.  The action was explicitly constructed in
$(2,2)$ superspace in \cite{CEGK}, to which we refer the reader
for a more detailed discussion.

It is convenient to define the coordinates $z^\pm = X^0 \pm X^1,
x= X^2 + i X^3$ and $y= X^4 + i X^5$.  The two-dimensional $(2,2)$
superspace is spanned by $(z^+,z^-,\theta^+, \theta^-,
\bar\theta^+, \bar\theta^-)$.  The four-dimensional fields
corresponding to D3-D3 strings are described by  $(2,2)$
superfields with extra continuous labels $x,\bar x$, while fields
associated to the D$3^{\prime}$-D$3^{\prime}$ strings have the
extra labels $y,\bar y$. Although the four-dimensional parts of
the action will look unusual in $(2,2)$ superspace,  this notation
makes sense since only a two-dimensional supersymmetry is
preserved.\footnote{The procedure of writing supersymmetric
$d$-dimensional theories in terms of a lower dimensional
superspace has been discussed in various places
\cite{HG,EGK,Hebecker,CEGK}. }  The fields associated with
D3-D$3^{\prime}$ strings are trapped at the intersection and have
no extra continuous label.

Let us first consider $S_{\rm D3}$, which involves $(2,2)$
superfields of the form $F(z^+,z^-,\theta, \bar\theta | x,\bar
x)$. The $(2,2)$ fields appearing in this action are a vector
superfield $V$, together with three adjoint chiral superfields
$Q_1, Q_2$ and $\Phi$. The gauge connections $A_{0,1}$ of the
$(2,2)$ vector multiplet and the complex scalar $\phi$ of the
$(2,2)$ chiral field $\Phi$ combine to give the four components of
the four-dimensional gauge connection. The scalar components of
$V, Q_1$ and $Q_2$ combine to give the six adjoint scalars of the
four-dimensional ${\cal N} =4$ theory. The field content of the
second D3-brane (D$3^{\prime})$ is identical to that of the first
D3-brane with the replacements $x \rightarrow y, V \rightarrow
{\cal V}, Q_i \rightarrow S_i$ and $\Phi \rightarrow \Upsilon$.
The fields corresponding to D3-D$3^{\prime}$ strings are the
chiral multiplets $B$ and $\tilde B$ in the $(N, \bar N^{\prime})$
and $(\bar N, N^{\prime})$ representations of the $SU(N) \times
SU(N^{\prime})$ gauge group. Together $B$ and $\tilde B$ form a
$(4,4)$ hypermultiplet.

For simplicity,  we just present the superpotential, $W = W_{D3} +
W_{D3^{\prime}} + W_{D3-D3^{\prime}}$, where
\begin{align}\ \label{bulkaction1}
&W_{\rm D3} =   \int d^2x  \epsilon_{ij} {\rm\, tr\,} Q_i
[\partial_{\bar x} + g \Phi, Q_j], \quad W_{\rm D3^{\prime}}= \int
d^2y  \epsilon_{ij} {\rm \,tr\,} S_i [\partial_{\bar y} +
g\Upsilon, S_j]  \quad, \nonumber \\
&W_{\rm D3- D3^{\prime}} =  \frac{ig}{{2}} {\,\rm tr}\left(
  B \tilde B Q_1 - \tilde B B S_1\right) \quad.
\end{align}
The Lorentz invariance of the four-dimensional parts of the action
$S_{D3}$ or $S_{D3}$ is not manifest in $(2,2)$ superspace, but
can seen upon integrating out auxiliary fields. Consider  the
Lorentz invariant kinetic term for the scalar component $q_1$ of
the chiral superfield $Q_1$, ${\rm tr}\,\partial_\mu \bar q_1
\partial^{\mu} q_1$ with $\mu = 0,1,2,3$.  This term arises from a
combination of a $(2,2)$ K$\ddot{{\rm a}}$hler term ${\rm tr}\,
\bar Q_1 Q_1$ and the superpotential term ${\rm tr}\, Q^1
\partial_z Q^2$.

On the classical Higgs branch the scalar components $b$ and
$\tilde b$ of the chiral fields $B$ and $\tilde B$ have non-zero
expectation values. The scalar components $s_2$ and $q_2$ of the
chiral superfields $S_2$ and $Q_2$ also have expectation values
given by the vanishing of the F-terms for $S^1$ and $Q^1$:
\begin{align}
\frac{\partial W}{\partial q_1} = \partial_{\bar x} q_2 -
g\delta^2(x) b \tilde b &=0 \,,\qquad \frac{\partial W}{\partial
s_1}= \partial_{\bar y} s_2 - g\delta^2(y) \tilde b b = 0\,.
\label{holomeq}\end{align} With the geometric identifications $q_2
\sim y/\alpha^{\prime}$ and $s_2 \sim x/\alpha^{\prime}$, the
solutions of these equations give rise to holomorphic curves of
the form $x y = c\alpha^{\prime}$, when $2\pi i c= g b\tilde b = g
\tilde b b$.

\section{Intersecting D3-branes at a $\CC^2/\ZZ_k$ orbifold}

The $(4,4)$ defect CFT which describes intersecting D3-branes in
flat space has an $SU(2)_L \times SU(2)_R$ R-symmetry.
Geometrically, this corresponds to a rotation of the four
coordinates transverse to all the D3-branes.  These four directions
can be described by two complex coordinates $u$ and $w$. We shall
now consider the intersecting D3-branes in the orbifold geometry
defined by $(u,w^*) \sim \exp(2\pi i/k)(u, w^*)$.

\begin{figure}[!ht]
\begin{center}
\includegraphics[height=18cm,clip=true,keepaspectratio=true]{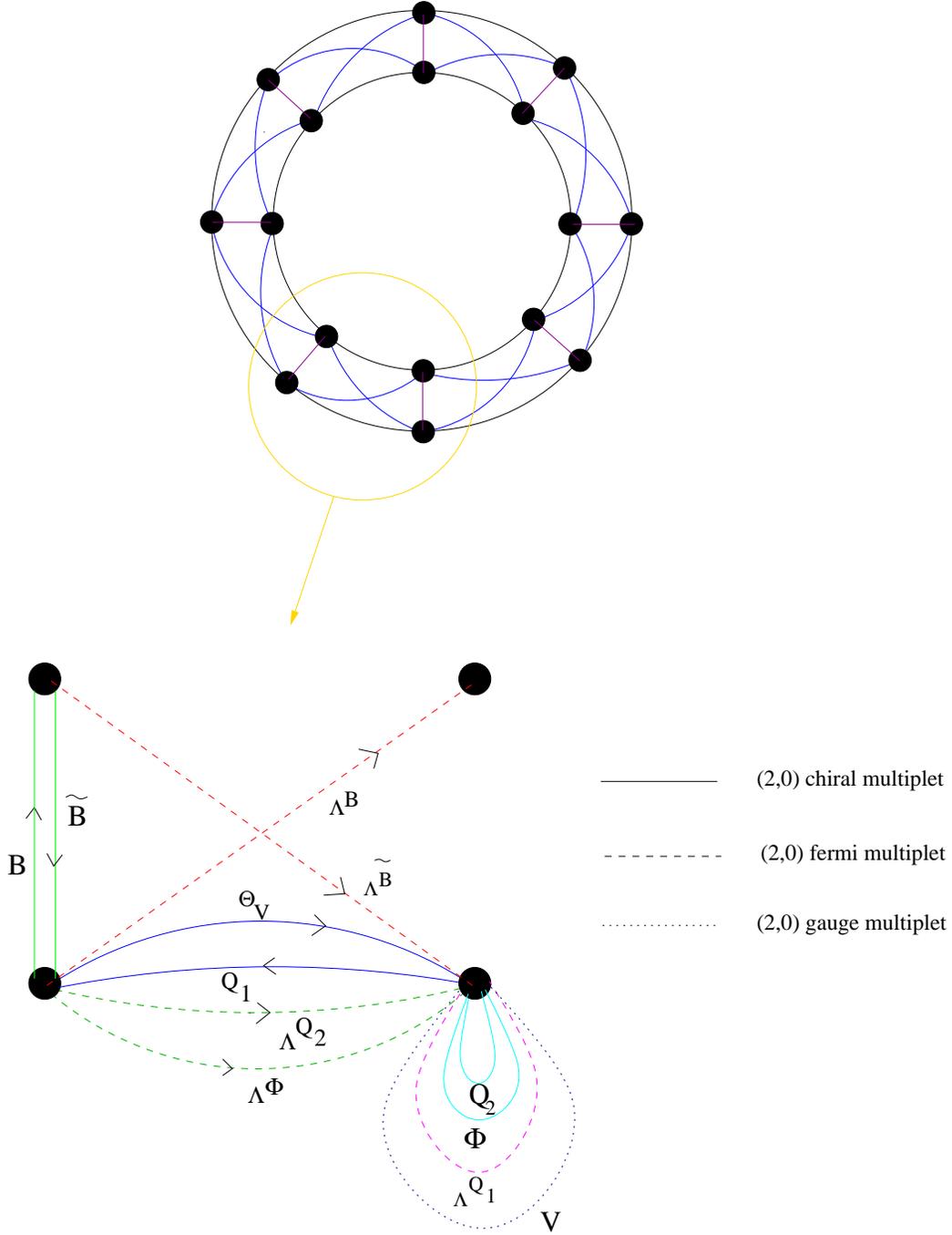}
\caption{Quiver diagram for intersecting D3-branes at a
$\CC^2/\ZZ_k$ orbifold (with $k$=8).  The nodes of the inner and
outer circle are associated with the $SU(N')^k$ and $SU(N)^k$
gauge groups respectively.  The parts which have not been drawn in
the detailed ``close-up'' are easily inferred from the $\ZZ_k$
symmetry and by swapping D3 degrees of freedom with D$3^{\prime}$
degrees of freedom.}\label{superquiver}
\end{center}
\end{figure}

The action in this background can be obtained from that of the
$(4,4)$ defect CFT with $U(Nk) \times U(N^{\prime}k)$ gauge group
by projecting out degrees of freedom which are not invariant under
the $\ZZ_k$ orbifold action.  The $\ZZ_k$ orbifold is embedded in
a combination of the gauge symmetry and the $SU(2)_R$ R-symmetry.
The result of the projection is a $(4,0)$ defect CFT, with $U(N)^k
\times U(N^{\prime})^k$ gauge group and an $SU(2)_L$ R-symmetry.
The field content and gauge transformation properties are
summarized by the quiver diagram in figure~\ref{superquiver}. The
inner and outer rings of the quiver diagram separately correspond
to the ${\cal N} =2, d=4$ superconformal Yang Mills theory which
was used in \cite{Arkani-Hamed} to (de)construct the six-dimensional
$(2,0)$ theory describing parallel M5-branes. The ``spokes'' 
stretching between the inner and outer rings correspond to
the degrees of freedom localized at the two dimensional intersection.
These degrees of freedom descend from the $(2,2)$ chiral
superfields $B$ and $\tilde B$ of the parent $(4,4)$ theory, and
consist of $(2,0)$ chiral multiplets $B_j$ and $\tilde B_j$ as
well as $(2,0)$ Fermi multiplets $\Lambda^B_{j,j+1}$ and
$\Lambda^{\tilde B}_{j,j+1}$.  The indices $j$ label nodes of the
quiver diagram and run from $1$ to $k$.  In an appropriate limit,
the ``spoke'' fields will be seen to correspond to the tensionless
strings of the M5-M5 intersection.

\section{Generating two extra dimensions}

In the $(4,0)$ defect CFT, two extra dimensions are generated at a
point on the Higgs branch where the bifundamental scalars of the
inner and outer rings of the quiver diagram have equal non-zero
expecation values. This can be seen directly from the analogue of
a $(2,0)$ superpotential for the $(4,0)$ defect CFT. We shall
focus on the part of the superpotential involving the degrees of
freedom at the intersection, which is given by
\begin{align}
W_{\rm D3-D3'} = g {\rm tr}_{N\times N}& \left(
\Lambda^B_{j,j+1}(\tilde B_{j +1}Q^1_{j,j+1} - S^1_{j+1,j} \tilde
B_j) + \Lambda^{Q^1}_j B_j\tilde B_j \right) \nonumber\\
+\ g{\rm tr}_{N^{\prime} \times N^{\prime}} & \left.\left(
\Lambda^{\tilde B}_{j,j+1} (Q^1_{j+1,j}B_j - B_{j+1} S^1_{j+1,j})
- \Lambda^{S^1}_j \tilde B_j B_j \right)\right\vert_{\bar\theta^+
= 0} \,. \label{superpotential}\end{align} In this expression the
$\Lambda$ fields are $(2,0)$ Fermi multiplets, while the other
fields are $(2,0)$ chiral multiplets. At the point in moduli space
for which extra dimensions are generated, $\langle
q^1_{j,j+1}\rangle = vI$ and $\langle s^1_{j,j+1}\rangle= vI$.
Expanding about this point in the moduli space, the quadratic part
of the superpotential is
\begin{align}\label{latkin}
W_{\rm D3-D3'} = gv\int d\theta^+ {\rm tr} \left. \left[
\Lambda^{\tilde B}_{j,j+1}(B_j - B_{j+1}) + (\tilde B_{j+1} -
\tilde B_j)\Lambda^B_{j,j+1}\right]\right\vert_{\bar\theta^+ = 0}
\, .
\end{align}
This superpotential can be viewed as a kinetic term in a compact
lattice direction, corresponding to the circular quiver diagram,
with $k$ sites and radius $R \sim \frac{k}{gv}$. Strictly
speaking, there are additional contributions to the kinetic term
arising from terms related to the superpotential by $(4,0)$
supersymmetry. These only change the radius by a numerical factor.
By a discrete ($\ZZ_k$) Fourier transform of modes localized at
the intersection, one obtains Kaluza-Klein modes on the extra
discrete circle.  There is yet another discrete circle due to the
S-duality of the D3-D3-orbifold system.  Under S-duality
$g\rightarrow k/g$,  such that the radius of the other circle is
$R_D = g/v$.  The continuum limit which keeps both radii fixed is
$k\rightarrow\infty$ with $g \sim \sqrt{k}$ and $v\sim \sqrt{k}$.

To determine whether the spoke degrees of freedom correspond to
strings or particles in four dimensions, it is helpful to deform
the theory by going to a point in the moduli space where they
become massive. This is accomplished by setting $Y_{j,j+1} =
(v+\Delta,v+\Delta)$ and $Y^{\prime}_{j,j+1}=(v-\Delta,
v-\Delta)$,  where $Y$ and $Y^{\prime}$ are the $SU(2)_L$ doublets
of bifundamental scalars on the inner and outer rings of the
quiver.  Note that $q^1_{j,j+1}$ belongs to the doublet
$Y_{j,j+1}$ while $s^1_{j,j+1}$ belongs to $Y^{\prime}_{j,j+1}$.
The real parameter $\Delta$ must be scaled as $1/\sqrt{k}$ to
consistently generate two extra dimensions,  but  still has an
effect as $k\rightarrow\infty$. The effective superpotential at
this point in moduli space is
\begin{align}
  W &= gv \int d\theta^+ {\rm tr} \left.\left[ B_j({\Lambda^{\tilde
          B}_-}_{j,j+1} - {\Lambda^{\tilde B}_-}_{j-1,j}) +
      {\Lambda^B_-}_{j,j+1}(\tilde B_{j+1} - \tilde
B_j)\right]\right\vert_{\bar \theta^+} \nonumber \\
  &+ \,g\Delta\int d\theta^+ {\rm tr} \left.\left[B_j({\Lambda^{\tilde
          B}_-}_{j,j+1} + {\Lambda^{\tilde B}_-}_{j-1,j}) +
      {\Lambda^B_-}_{j,j+1}(\tilde B_{j+1} + \tilde
      B_j)\right]\right\vert_{\bar \theta^+} \label{massterm} \,.
\end{align}
For large $k$ and fixed lattice momentum $n$, diagonalizing the
mass matrix for the fundamental spoke degrees of freedom gives
\begin{align}
M^2 = (g\Delta)^2 + (n/R)^2\,,
\end{align}
where $n$ is the lattice momentum.  For simplicity let us set
$n=0$, so that $m = g\Delta$. The S-dual modes then have
$m_D=\frac{k}{g}\Delta$. Since $m/m_D = R_D/R$, the fundamental
spoke degrees of freedom should be interpreted as strings wrapping
the cycle of radius $R_D$,  while their S-duals wrap the cycle of
radius $R$. The string tension is $T = \frac{m}{2\pi R_D} =
\frac{m_D}{2\pi R} = v\Delta$.  As $\Delta \rightarrow 0$, we
obtain a theory of tensionless strings.

\section{Matching R-symmetries and a 't Hooft Anomaly}
The M5-M5 intersection has ${\cal N} =2, d=4$ supersymmetry with
$SU(2) \times U(1)$ R-symmetry.  We shall now argue that the
$SU(2)$ R-symmetry should be identified with the $SU(2)_L$
R-symmetry of the $(4,0)$ defect CFT.  This is not a trivial
identification, since the $SU(2)_L$ R-symmetry is apparently
spontaneously broken at the point in moduli space required to
generate two extra dimensions, whereas the $SU(2)$ R-symmetry of
the M5-M5 intersections is only spontaneously broken when the
M5-branes are transversely separated, corresponding to a non-zero
string tension $T = |\vec X - \vec X^{\prime}|$.  The quantity
$\vec X - \vec X^{\prime}$ is a triplet under the $SU(2)$
R-symmetry.  Note that the string tension in the $(4,0)$
description of M5-M5 intersection can be written in an $SU(2)_L$
invariant way as $T = |Y^{\dagger}\vec\sigma Y - Y^{\prime\dagger}
\vec\sigma Y^{\prime}|$.  This suggests the identification of
$Y^{\dagger}\vec\sigma Y - Y^{\prime\dagger} \vec\sigma
Y^{\prime}$ with $\vec X - \vec X^{\prime}$, which in turn
requires the identification of $SU(2)_L$ with the $SU(2)$
R-symmetry of the M5-M5 system. This suggests that, for vanishing
string tension, $SU(2)_L$ is unbroken as far as the non-trivial
dynamics is concerned.

The only two-dimensional degrees of freedom of the $(4,0)$ defect
CFT which are charged under $SU(2)_L$ are doublets of negative
chirality.  Consequently there is a 't Hooft anomaly in $SU(2)_L$.
Note that such anomalies are protected quantities which are
readily computable in the strong coupling limit.  It is remarkable
that the theory which (de)constructs the M5-M5 intersection is
chiral,  and it would be surprising if this chirality did not have
consequences in the continuum limit. Assuming that the 't Hooft
anomaly we have just discovered survives in this limit,  it should
be interpreted as an $SU(2)$ anomaly due to tensionless strings.
While there are no local $SU(2)$ anomalies in four-dimensional
quantum field theories, we know of nothing which excludes the
possibility for four-dimensional theories of tensionless strings.
Unfortunately,  we can not yet state conclusively that the
$SU(2)_L$ anomaly has a non-zero continuum limit.

Assuming that this anomaly exists in the continuum limit, it is
somewhat similar in spirit to the known $Spin(5)$ 't Hooft anomaly
of the $(2,0)$ theory of parallel M5-branes
\cite{Witteneff,FHMM,HMM,bonorarinaldi,LMT,BHR,intriligator}. The
$Spin(5)$ R-symmetry corresponds to the Lorentz symmetry of the
five directions transverse the M5-branes. This symmetry is gauged
by the coupling to gravity.  It turns out there there is an
anomaly in diffeomorphisms of the normal bundle due to the long
wavelength Chern-Simons terms of eleven dimensional supergravity
in the presence of magnetic (M5-brane) sources.  Consistency
requires that this anomaly is cancelled by a contribution due to
the degrees of freedom propagating on the M5-brane.  This gives an
indirect derivation of the t' Hooft anomaly for the
six-dimensional $(2,0)$ theory.  At present there is no direct
derivation in the non-abelian case.  In principle, it should be
possible to reproduce this anomaly from the (de)constructed
description of the six-dimensional $(2,0)$ theory
\cite{Arkani-Hamed}. Unfortunately this is complicated by the fact
that the $Spin(5)$ R-symmetry is not manifest in this description,
and is realized only in the continuum limit.

If there is an R-symmetry anomaly of the M5-M5 intersection due to
tensionless strings,  it should also be computable via the
assumption of anomaly cancellation in M-theory. The question of
the existence of this anomaly is worth pursuing further,
particularly in light of its relation to black hole entropy
\cite{HMM}.

{\bf Acknowledgement} The authors wish to thank G. Cardoso, A.
Hanany, R. Helling, B. Ovrut, S. Ramgoolam, W.~Skiba, D.~Tong and
J.~Troost for helpful discussions. The research of J.E., Z.G.~and
I.K.~is funded by the DFG (Deutsche Forschungsgemeinschaft) within
the Emmy Noether programme, grant ER301/1-2. N.R.C.\ is supported
by the DOE under grant DF-FC02-94ER40818, the NSF under grant
PHY-0096515 and NSERC of Canada.


\end{document}